\RequirePackage{fix-cm}

\documentclass[twocolumn,epjc3]{svjour3}  
\smartqed  
\RequirePackage{graphicx}

\RequirePackage{float}
\RequirePackage{amsmath}
\RequirePackage{lineno}
\RequirePackage[colorlinks,citecolor=blue,urlcolor=blue,linkcolor=blue]{hyperref}
\journalname{Preprint}

\modulolinenumbers[5] 

\begin{document}

\title{Possible evidence for the production of Ar$_2^{*-}$ metastable negative molecular ions in gaseous argon of two-phase detectors for dark matter searches 
	}


\author{A.~Buzulutskov\thanksref{addr1,addr2}
        \and 
        E.~Frolov\thanksref{addr1,addr2,e1}
		\and
		E.~Borisova\thanksref{addr1,addr2}
		\and
		V.~Nosov\thanksref{addr1,addr2}
		\and
		V.~Oleynikov\thanksref{addr1,addr2}
		\and
		A.~Sokolov\thanksref{addr1,addr2}
        }

\thankstext{e1}{geffdroid@gmail.com (corresponding author)}


\institute{Budker Institute of Nuclear Physics SB RAS, Lavrentiev avenue 11, 630090 Novosibirsk, Russia \label{addr1} 
\and Novosibirsk State University, Pirogova street 2, 630090 Novosibirsk, Russia \label{addr2}
}

\date{Received: date / Accepted: date}

\maketitle

\begin{abstract}
\sloppy Our recent studies of electroluminescence (EL) properties in two-phase argon detectors for dark matter searches have revealed the presence of unusual delayed pulses in the EL signal in the form of two slow components with time constants of about 5 and 50 $\mu$s. 
These components were shown to be present in the charge signal itself, which clearly indicates that drifting electrons are temporarily trapped on two states of metastable negative argon ions which have never been observed before. In this work, using the pressure dependence of the ratio of slow component contributions measured in experiment, it is deduced that these states are those of two types of metastable negative molecular ions, $\mathrm{Ar}_2^{*-}(b \ ^4\Sigma_u^-)$ and  $\mathrm{Ar}_2^{*-}(a \  ^4\Sigma_g^+)$ for the higher and lower energy level respectively.

\end{abstract}


\section{Introduction}\label{intro}

Two-phase (liquid-gas) Ar detectors with electroluminescence (EL) gap are commonly used in large-scale dark matter particle (WIMP) search and neutrino detection experiments \cite{Aalseth18,Akimov21}. In two-phase detectors, the EL effect~\cite{Buzulutskov20} is used in the gas phase to record the primary ionization signal (S2), delayed with respect to that of primary scintillation (S1). At low-mass WIMP searches~\cite{Agnes18,Aprile20}, the background rate on a long time frame (at ms scale), defined by pulses delayed with respect to the main S2 signal, becomes of primary importance. Such pulses were observed both in Xe-based~\cite{Aprile22} and Ar-based~\cite{Agnes23} detectors and were interpreted as being produced by electrons trapped in the liquid bulk or at the liquid-gas interface.

Our recent studies of EL properties in two-phase Ar detectors~\cite{Bondar20,Buzulutskov22} have revealed the presence of other additional unusual delayed pulses in the EL signal in the form of two slow components (in addition to fast component corresponding to no delay) with time constants of about 5 and 50~$\mu$s, called ``slow'' and ``long'' component respectively. They emerged at higher reduced electric fields in the EL gap, above the threshold of 5~Td which is 1~Td larger than that of excimer EL. Moreover, they were shown to be present in the charge signal itself, induced in the EL gap~\cite{Buzulutskov23a}. The latter fact indicates that some of the electrons are temporarily trapped during their drift in the EL gap on two new metastable states of negative Ar ions of yet unknown nature.

In this work, a non-trivial pressure dependence of the slow to long component ratio is observed, which helps to unravel the slow component puzzle: it is inferred that the two states responsible for the slow components are those of two types of metastable negative molecular ions, Ar$_2^{*-}$, which have never been observed before.


\section{Results}\label{results}

Experimental setup used in this work is described elsewhere~\cite{Buzulutskov22}.
The detector was a two-phase time projection chamber (TPC), composed of drift region and electron emission region in the liquid phase and EL gap in the gas phase. 
It was operated in the equilibrium state at four saturated vapor pressures, of 0.75, 1.00, 1.25 and 1.50~atm, corresponding to temperatures of 84.7, 87.3, 89.5 and 91.3~K~\cite{Stewart89}. The EL signal was recorded using four cryogenic PMTs and 5$\times$5 SiPM matrix. Due to lack of any wavelength shifter (WLS) the devices were sensitive to EL in the visible and NIR range, provided by the neutral bremsstrahlung (NBrS) EL mechanism~\cite{Buzulutskov18,Borisova21,Borisova22,Henriques22,Milstein22}.

\begin{figure}[!t]
	\centering
	\includegraphics[width=1.0\linewidth]{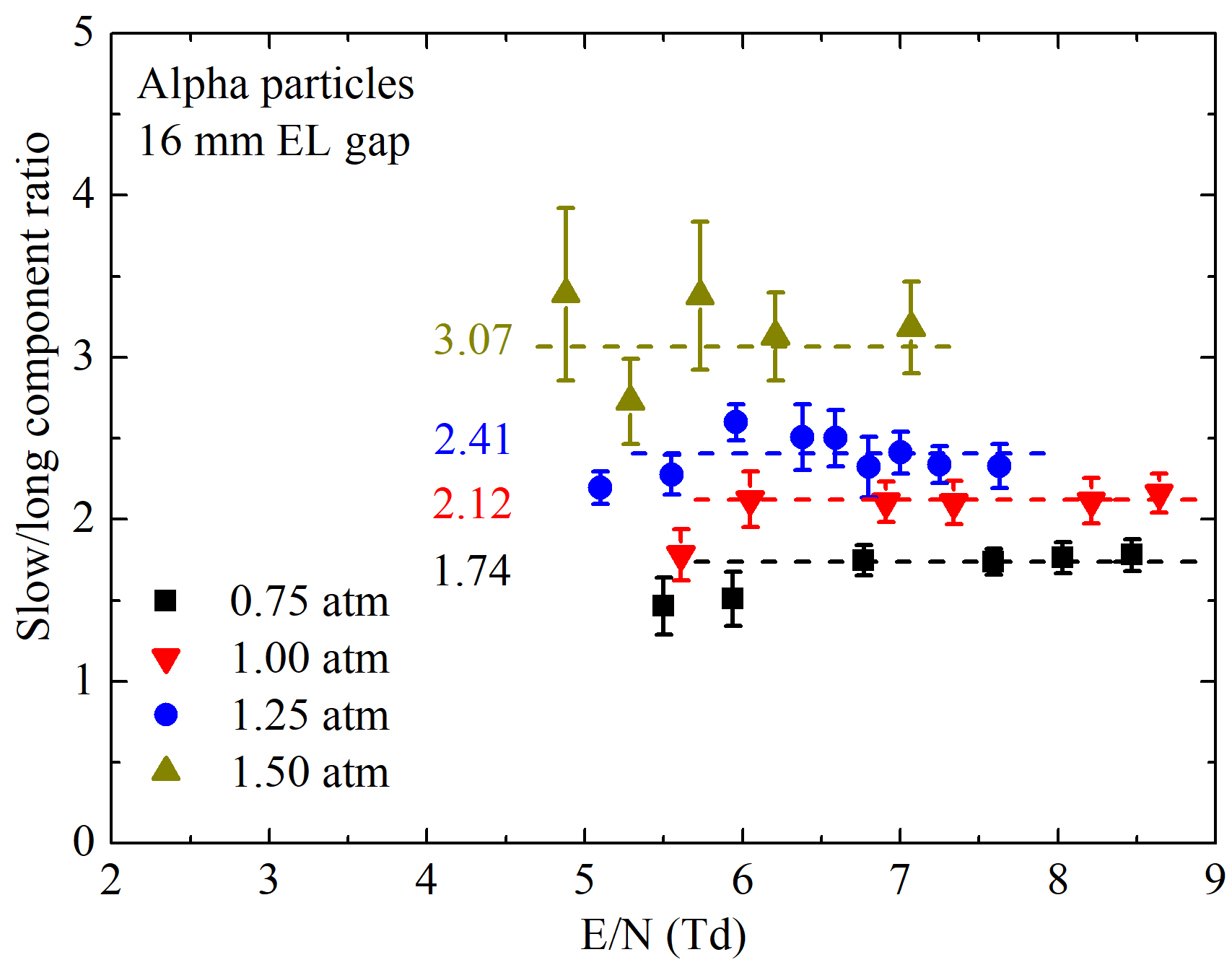}
	\caption{Ratio of slow to long component contribution to the overall EL signal of two-phase Ar detector as a function of the reduced electric field at different gas pressures. The EL signals were produced by the NBrS mechanism in the visible and NIR range and were obtained with the $^{238}$Pu alpha-particle source and 16~mm EL gap thickness. Dashed lines show average ratio values.
	}
	\vspace{-10pt}
	\label{fig:ratio_detailed}
\end{figure}

The experimental results of this work are presented in Fig.~\ref{fig:ratio_detailed}, showing the contribution to the overall EL signal for the slow component divided by the same contribution for the long component (further referred to as ratio of slow to long component contribution) for different electric fields and pressures. These results confirm our previous observation~\cite{Buzulutskov22} that this ratio is independent of the electric field, indicating that slow and long components have the same production mechanism. In particular, $\chi^{2}/ndf$ values for constant fits are 0.54, 0.04, 0.80 and 0.69 for 0.75, 1.0, 1.25 and 1.5 atm pressure respectively. Accordingly, the linear slope is close to zero for all pressures, namely below 0.06 Td$^{-1}$ with 95\% confidence level. On the other hand, the value of this ratio depends on the pressure: it substantially increases with pressure from 1.7 at 0.75 atm to 3.1 at 1.5 atm. 

Before moving on, let us summarize the puzzling properties of the slow components revealed in~\cite{Bondar20,Buzulutskov22,Buzulutskov23a} and this work:
\begin{enumerate}
\item There are two slow components in proportional EL and corresponding charge signal, with time constants of about 4-5~$\mu$s and 50~$\mu$s.

\item Both slow components emerge at a threshold in reduced electric field of about 5~Td which is 1~Td above the onset of excimer EL.

\item The ratio of slow to long component contribution, being independent of the electric field, substantially increases with pressure.

\item The slow components contributions and time constants increase with the electric field.

\item The slow components contributions decrease with temperature, practically disappearing at room temperature.

\end{enumerate}

Below we present a model for the formation of the slow components that explains all their puzzling properties. In this model, it is proposed that the formation of two slow components is provided by two parent states of excited atoms, two daughter states of negative atomic ions and two granddaughter states of negative molecular ions, the latter being directly responsible for the production of slow and long components by trapping drifting electrons, thus explaining the 1st property.

\begin{figure*}[!t]
	\centering
	\includegraphics[width=0.75\linewidth]{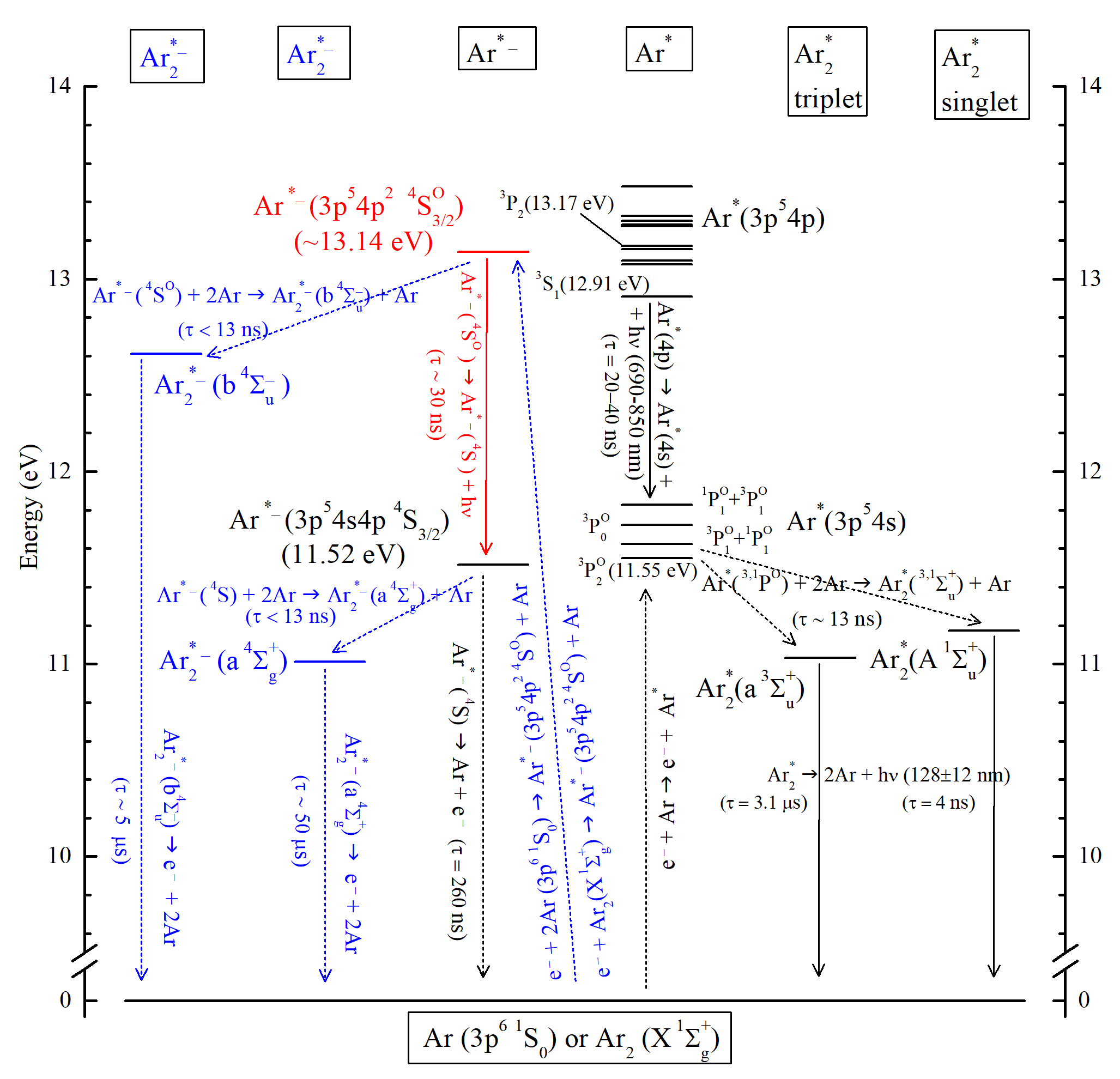}
	\caption{Energy levels and production reactions for atomic excited states Ar$^*$~\cite{NIST,Buzulutskov17}, negative atomic ion states Ar$^{*-}$ \cite{Bunge82,Bae1985:Ar_ion_state,Ben-Itzhak1988:Ar_ion_state,Pedersen98}, molecular excited states Ar$_2^{\ast}$ (minimums of potential energy curves~\cite{Yates83,Duplaa96}) and negative molecular ion states Ar$_2^{\ast-}$ proposed in this work (minimums of potential energy curves), all given in LS-coupling notation. The atomic levels are shown with their mixing determined from~\cite{Katsonis11}. Black solid arrows indicate the radiative transitions observed in experiments: Ar$_2^{\ast}$ in the VUV~\cite{Cheshnovsky72} and Ar$^{\ast}$ in the NIR~\cite{Lindblom88}. Time constants for the VUV and the NIR transitions are taken from~\cite{Buzulutskov17} and~\cite{Schulze08} respectively. Red solid arrow shows predicted dipole radiative transition of Ar$^{*-}(3p^54p^2 \ ^4S^{O}_{3/2})$~\cite{Bunge82} similar to its parent Ar$^{*}(3p^54p)$. Dashed arrows indicate the non-radiative transitions with their time constants reduced to the gas density at 1.0 atm pressure and temperature of 87 K. The ground states are composed of argon atoms $\mathrm{Ar}\,(3p^{6}\, ^1S_0)$ and Van-der-Waals molecules $\mathrm{Ar}_2(\mathrm{X}\, ^1\Sigma^{+}_{g})$. The levels and reactions in black, red and blue are those observed in experiment, those theoretically predicted elsewhere and those proposed in this work respectively.}
	\vspace{-10pt}
	\label{fig:levels}
\end{figure*}

Accordingly, Fig.~\ref{fig:levels} shows the energy levels and production reactions for atomic excited states Ar$^*$~\cite{NIST,Buzulutskov17}, negative ion states Ar$^{*-}$~\cite{Bunge82,Bae1985:Ar_ion_state,Ben-Itzhak1988:Ar_ion_state}, molecular excited states Ar$_2^{\ast}$ ~\cite{Yates83,Duplaa96} and negative molecular ion states Ar$_2^{\ast-}$, the latter being proposed in this work.

The full chain of appropriate reactions, responsible for the production of slow and long components, is given below:
\begin{eqnarray}
\label{Neg-Ion-form-1}
e^- + 2\mathrm{Ar}(3p^6 \ ^1S_0) \rightarrow  \mathrm{Ar}^{*-}(3p^54p^2 \ ^4S_{3/2}^O) + \mathrm{Ar}\; , \\
\label{Neg-Ion-form-2}
e^- + \mathrm{Ar}_2(\mathrm{X}\, ^1\Sigma^{+}_{g}) \rightarrow  \mathrm{Ar}^{*-}(3p^54p^2 \ ^4S_{3/2}^O) + \mathrm{Ar}\; , \\
\label{Neg-Ion-form-3}
\mathrm{Ar}^{*-}(3p^54p^2 \ ^4S_{3/2}^O) + 2\mathrm{Ar} \rightarrow  \mathrm{Ar}_2^{*-}(b \ ^4\Sigma_u^-) + \mathrm{Ar}\; , \\
\label{Neg-Ion-form-4}
\mathrm{Ar}^{*-}(3p^54p^2 \ ^4S_{3/2}^O) \rightarrow  \mathrm{Ar}^{*-}(3p^54s4p \ ^4S_{3/2}) + h\nu\; , \\
\label{Neg-Ion-form-5}
\mathrm{Ar}^{*-}(3p^54s4p \ ^4S_{3/2}) + 2\mathrm{Ar} \rightarrow  \mathrm{Ar}_2^{*-}(a \ ^4\Sigma_g^+) + \mathrm{Ar}\; , \\
\label{Neg-Ion-form-6}
\mathrm{Ar}_2^{*-}(b \ ^4\Sigma_u^-) \rightarrow  e^- + 2\mathrm{Ar}\; , \\
\label{Neg-Ion-form-7}
\mathrm{Ar}_2^{*-}(a \ ^4\Sigma_g^+) \rightarrow  e^- + 2\mathrm{Ar}\; .
\end{eqnarray}
$\mathrm{Ar}_2(\mathrm{X}\, ^1\Sigma^{+}_{g})$ in reaction \ref{Neg-Ion-form-2} are Van-der-Waals molecules \cite{Buzulutskov22,Buzulutskov17,Smirnov84}, providing 2.7\% of the ground states in the gas phase at 87 K. Their contribution to the formation of slow components was introduced in~\cite{Buzulutskov22} to explain the temperature dependence (the 5th property).

\sloppy Regarding negative atomic ion states, both $\mathrm{Ar}^{*-}(3p^54s4p \ ^4S_{3/2})$ and $\mathrm{Ar}^{*-}(3p^54p^2 \ ^4S^{O}_{3/2})$ states were predicted in \cite{Bunge82} by calculation of their electron affinity. They are composed by adding $4p$ electron to the parent neutral excitation state, $\mathrm{Ar}^{*}(3p^54s \ ^3P^{O}_2)$ and $\mathrm{Ar}^{*}(3p^54p \ ^3P_2)$ respectively, thus defining their quantum numbers according to selection rules for angular momentum and parity conservation \cite{Elyashevich01}. The lower energy negative ion state $\mathrm{Ar}^{*-}(3p^54s4p \ ^4S_{3/2})$ was observed in atomic beam experiments \cite{Bae1985:Ar_ion_state,Ben-Itzhak1988:Ar_ion_state,Pedersen98} with a life time of 260~ns~\cite{Ben-Itzhak1988:Ar_ion_state} and binding energy (electron affinity) of 32.5~meV with respect to the parent state $\mathrm{Ar}^{*}(3p^54s \ ^3P^{O}_2)$~\cite{Pedersen98}. 
Following~\cite{Bunge82}, we may assign the close value of electron affinity (of about 30~meV) to the higher energy negative ion state $\mathrm{Ar}^{*-}(3p^54p^2 \ ^4S^{O}_{3/2})$ with respect to its parent state $\mathrm{Ar}^{*}(3p^54p \ ^3P_2)$.

The lower energy state of negative atomic ion has even parity and is metastable while the higher state has odd parity and is linked to the lower one by electrical dipole transition~\cite{Bunge82} (reaction~\ref{Neg-Ion-form-4}). As it is similar to the transition between the parent states, $\mathrm{Ar}^{*}(3p^54p) \rightarrow \mathrm{Ar}^{*}(3p^54s) + h\nu$, it should have photon emission in the NIR with similar time constant, of about 30~ns.

Despite its energy favor, the production of lower energy state of negative atomic ion  $\mathrm{Ar}^{*-}(3p^54s4p \ ^4S_{3/2})$ in electron-atom collisions in reactions similar to that of~\ref{Neg-Ion-form-1} and~\ref{Neg-Ion-form-2} is forbidden due to parity conservation. Indeed, the total angular momentum of both negative atomic ion states is 3/2, meaning that the electron in initial $e^- +  \mathrm{Ar}(3p^6 \ ^1S_0)$ system should be in $p$-wave with orbital momentum $l=1$ and odd parity (equal to $(-1)^l$). This means that the resulting negative atomic ion state should also have odd parity. The higher energy state $\mathrm{Ar}^{*-}(3p^54p^2 \ ^4S^{O}_{3/2})$ satisfying this condition have the same energy level as those of $\mathrm{Ar}^{*}(3p^54p)$ configuration and therefore the same production threshold in reduced electric field, of 5~Td \cite{Buzulutskov22}, explaining the 2nd property of the slow components. Note that the lower energy state of negative atomic ions $\mathrm{Ar}^{*-}(3p^54s4p \ ^4S_{3/2})$ is populated by radiative decay of the upper energy state (reaction~\ref{Neg-Ion-form-4}).

One can see that the upper energy state of negative atomic ions $\mathrm{Ar}^{*-}(3p^54p^2 \ ^4S^{O}_{3/2})$ is a branch point of two competing reactions, \ref{Neg-Ion-form-3} and \ref{Neg-Ion-form-4}, leading to the production of respectively higher and lower energy states of negative molecular ions, $\mathrm{Ar}_2^{*-}(b \ ^4\Sigma_u^-)$ and  $\mathrm{Ar}_2^{*-}(a \ ^4\Sigma_g^+)$. Their quantum numbers result from given states of an atom and atomic ion according to selection rules (see section 24.3 in~\cite{Elyashevich01}). The higher negative molecular ion state is produced in reaction~\ref{Neg-Ion-form-3} of three-body collision (with a time constant below 13~ns~\cite{Buzulutskov17}) and thus strongly depends on atomic density (pressure), in ideal case as $N^2$. The lower negative molecular ion state is produced in competing reaction \ref{Neg-Ion-form-4} and successive reaction \ref{Neg-Ion-form-5}, the former having comparable time constant (30~ns) and being independent of the atomic density. Accordingly, the ratio of population of the higher and lower energy negative molecular ion states, responsible for the slow and long component respectively, should substantially increase with atomic density (pressure). This fact explains the 3rd property.

Finally, the 4th property is explained by rates of reactions \ref{Neg-Ion-form-1} and \ref{Neg-Ion-form-2} increasing with electric field due to increase in the number of electrons with energy exceeding the reaction threshold.

\section{Conclusions}

In this work, all puzzling properties of slow components of electroluminescence (EL) signal in two-phase Ar detectors have been successfully explained by the hypothesis that drifting electrons in the EL gap are temporarily trapped on two new metastable states of negative Ar ions. Using the pressure dependence of the ratio of slow to long component contribution measured in experiment, it is deduced that these new states are those of two metastable negative molecular ions, $\mathrm{Ar}_2^{*-}(b \ ^4\Sigma_u^-)$ and  $\mathrm{Ar}_2^{*-}(a \  ^4\Sigma_g^+)$ for the higher and lower energy level respectively. 
They have corresponding decay times of around 5 and 50 $\mu$s. 

The discovery of new kinds of negative Ar ions may pave the way for understanding the background in low-mass WIMP searches and for the development of neutral atomic Ar beams for plasma physics and space thrusters.

\section*{Acknowledgments}

This work was supported in part by Russian Science Foundation (project no. 20-12-00008, \url{https://rscf.ru/project/20-12-00008/}).


\bibliographystyle{spphys_modified}       
\bibliography{Manuscript}   

%
%

\end{document}